# Mining high on-shelf utility itemsets with negative values from dynamic updated database

Ms. Anjali N. Radkar
M.E. Student, Department of Computer Engineering,
D.Y Patil College of Engineering, Akurdi,
Savitribai Phule Pune University, Pune, India.
anjali_radkar@yahoo.com

Ms. S.S. Pawar
Asst. Prof., Department of Computer Engineering,
D.Y. Patil College of Engineering, Akurdi,
Savitribai Phule Pune University, Pune, India.
psoudamini@yahoo.co.in

*Abstract*— **Utility mining emerged to overcome the limitations of frequent itemset mining by considering the utility of an item. Utility of an item is based on user's interest or preference. Recently, temporal data mining has become a core technical data processing technique to deal with changing data**. **On-shelf utility mining considers on-shelf time period of item and gets the accurate utility values of itemsets in temporal database. In traditional on-shelf utility mining, profits of all items in databases are considered as positive values. However, in real applications, some items may have negative profit. Most of the traditional algorithms are used to handle static database. In practical situations, temporal databases are continually appended or updated. High on-shelf utility itemsets needs to be updated. Re-running the temporal mining algorithm every time is ineffective since it neglects previously discovered itemsets. It repeats the work done previously. In this paper, an effective algorithm is proposed to find high on-shelf utility itemsets with negative values from the dynamic updated temporal database.**

I. INTRODUCTION

**Data mining** is the process of extracting interesting (previously unknown and potentially useful) information or patterns from large information repositories. Data mining task includes finding association rules, classification rules, clustering rules. Among them, mining association rules is the most popular task in data mining. It has two phases. In first phase, it finds the frequent itemsets based on a user-defined minimum support threshold. In second phase, it generates the association rules from the discovered frequent itemsets based on the user-defined minimum confidence threshold.

In association-rule mining, the frequent itemsets consider only the frequency of an item in a database. The relative importance such as price, weight or profit of an item inside a transaction is not considered. However, in real world business, some items or itemsets with low support in the data set may bring high profits due to their high price or high frequency within transactions. Such useful, profitable itemsets are missed by frequent itemset mining [1].

In Weighted Frequent itemset mining, weights of items such as unit profits of items in the databases are considered. If items appear infrequently, they might still be found if they have high weights. But in this framework, the quantities of items are ignored. Therefore it cannot satisfy the requirements of users who are interested in finding the itemsets with consideration of both quantity and profit [6].

Recently, **Utility itemset mining** [7] has been proposed to eliminate the limitation of frequent and weighted itemset mining. It considers utility of an item which is based on interesting measures like user's preference or interest. Utility mining measures the importance of an item. Thus utility mining is useful in real world market data. Utility of an item in a database is the product of external and local transaction utility. The local transaction utility and the external utility are usually respectively defined as quantity and profit in utility mining. The utility of an itemset is considered as the product of quantity and profit. If utility of an itemset is greater than or equal to predefined minimum utlity threshold, then this itemset is considered as high utility itemset. In real world applications, the utility value of an itemset can be profit, measure of design, page-rank, popularity or some other measures of user's preference. It is also useful in web click streams where importance of each website is different. Similarly, it is also useful in applications areas like web server logs, telecom call records.

Temporal data mining [4, 5] has attracted a lot of attention due to its practicality. Temporal data exist extensively in economics, finance, communication, and other areas such as weather forecasting. Temporal transaction database is divided into several partitions according to time periods. In specific time or season, some items or itemsets may have high frequency. Thus mining time-related knowledge is interesting and useful. In real-world applications, products in a store can be put on shelf and taken off multiple times. To indentify such itemsets, **on-shelf utility mining** is proposed [11] in which an on-shelf period of products is calculated to get more accurate utility values of itemsets in temporal databases. In practical situations, utility of some items may have **negative profit** or it is difficult to calculate its profit like free products. But combination of such products with positive profit products may give profit to the business. In practical business applications, temporal databases are **dynamic**. They are continually appended or updated. Thus the discovered high utility itemsets need to be updated.

The remainder of the paper is organized as follows. In section II, related work is described. In section III, problem statement and definitions are explained. In section IV, proposed algorithm for finding high on-shelf utility items of both positive and negative values from dynamic updated databases is described. In section IV, experimental results are shown.

## II. RELATED WORK

In this section, studies related to temporal association-rule mining, utility mining, on-shelf utility mining and FUP (Fast Updated Pattern) concept are briefly reviewed.

### A. Temporal Association Rule Mining

Association-rule mining is an important task in the data mining. Agrawal et al. [1] proposed several mining algorithms to find association rules from transaction data. These algorithms generate unnecessary candidate itemsets which increase memory wastage as well as execution time. To avoid these candidate itemsets, Han, Pei and Yin [2] have proposed FP- tree structure for storing frequent patterns and FP-growth algorithm for mining frequent itemsets. In real world applications, transactions are stored in database along with its occurring time. Tanbeer, Ahmad, Jeong, Lee [3] proposed CP-tree (compact pattern tree) which extracts database in one scan. The CP-tree is a dynamic tree which produces a highly compact frequency-descending tree structure at runtime.

Temporal data mining is proposed to find temporal patterns from a set of data with time. Ale et al. [4] proposed a mining approach for finding temporal association rules from a temporal transaction database. They considered transaction periods of products but ignored exhibition periods of the products. Chang et al. [5] proposed an algorithm which considers common exhibition period of a product combination in a store. It is an interval between first and last on-shelf time periods. In their approach, they have not considered the individual on-shelf time periods. All these approaches ignore that the product might be put on and taken off multiple times from the shelf in the store.

### B. High Utility Itemset Mining

Traditional association rule mining considers only the quantity i.e. the presence or absence of items in the transaction. However in reality, items are associated with both quantity and profit. In some situations, items with high profit may have low frequency in the transaction. Such products are neglected in traditional frequent itemset mining due to its low frequency. Thus Chan, Yang and Shen [7] proposed utility mining which not only considers quantities of products (items) but also their profits in a set of transactions to extract the itemsets. The downward-closure property in association rule mining cannot be maintained in utility mining due to its utility function. To solve downward closure problem, Liu, Liao and Choudhary [8] proposed Two-Phase algorithm to reduce number of candidate itemset to discover high utility itemsets. In first phase, they have used transaction weighted downward closure property called the transaction-weighted utilization (TWU) model to generate candidate itemset. Transaction weighted utility of an itemset is the sum of transaction utility of transaction which contains an itemset. In second phase, database is scanned once to find high utility itemsets.

V.S. Tseng et al. [9] proposed a novel method called THUI (Temporal High Utility Itemsets) to discover temporal high utility itemsets from data stream. In this method, few temporal high utility itemsets are identified by generating less high transaction weighted utilization 2-itemsets to reduce execution time. Li, Yeh and Chang [10] proposed the Isolated Items Discarding Strategy (IIDS). In this approach, isolated items are identified from transactions and these are ignored in the process of candidate itemset generation.

### C. On-shelf Utility Itemset Mining

Traditional utility mining ignores the on-shelf time periods of products in stores. To solve this problem, Lan and Tseng [11] proposed a new topic named on-shelf utility mining, which considered the on-shelf periods of items with the quantities and profits of items. Lan, Hong et al. [12] proposed two-phased mining algorithm to discover high on-shelf utility itemsets efficiently. In the first phase, the possible candidate on-shelf utility itemsets within each time period are extracted. In the second phase, the candidate on-shelf utility itemsets is checked for their actual utility values.

Chu et al. [13] proposed utility mining with negative item profit. They developed a two-phase algorithm for finding high-utility itemsets with negative profits from a transaction database. Li et al. [14] proposed two algorithms MHUI-BIT (Mining High-Utility Itemsets based on BITvector) and MHUI-TID (Mining High-Utility Itemsets based on TIDlist) for finding high utility itemsets from data streams. They have used Bitvector and TIDlist to improve performance of utility mining. They have also developed MHUI-BIT-NIP (MHUI-BIT with Negative Item Profits) and MHUI-TID-NIP (MHUI-TID with Negative Item Profits) for discovering itemsets with negative profit over continuous data stream.

To find negative itemsets by considering on-shelf period, Lan et al. [15] proposed an algorithm HOUN (High On-shelf Utility mining with Negative item values) to find the on-shelf high utility items which have both positive as well as negative profit. This algorithm has three database scan. In first scan, preprocessing is done by transforming database into corresponding time period and discover high periodical utility upper bound 2-itemsets. In second scan, high periodical utility itemsets are discovered and in third scan, it discovers high on-shelf utility itemsets with negative values. Radkar and Pawar [16] have made a survey of various itemset mining algorithms in which they have explained various approaches and techniques of itemset mining.

*D. FUP Conecpt*

Most of the traditional algorithms are used to handle static database. In real world applications, database is dynamic in nature. Transactions can be inserted, deleted or updated dynamically from database. Due to this, association rules must be re-evaluated. Some new items can be added or some old ones may be removed. Hong, Lin, Wu [17] proposed a fast updated FP-Tree (FUFP-tree) which gets updates when new transactions are inserted. It partitions items in four parts to decide whether they are large or small in original and modified database. Based on that, tree is updated either by removing or by adding nodes to it. Ahmed, Tanbeer, Jeong, Lee [18] proposed the tree structure to perform incremental mining of high utility itemsets. They proposed IHUP (Incremental High Utility Pattern) Lexicographic tree to arrange items in lexicographic order and IHUP Transaction tree. Gharib et al. [19] proposed an incremental algorithm to maintain temporal association rules. They developed a technique to updated previously generated candidate itemsets to avoid re-running of whole process of itemset mining.

Lan, Lin, Hong, and V. S. Tseng [20] proposed an algorithm for average utility mining to update average utility itemsets in dynamic databases. When records are modified from the original database, itemsets are partitioned into four parts according to whether they are large or small in the original database and whether their count difference is positive or negative in the modified records. Each part is then processed to maintain the discovered knowledge in its own way. Lin, Zhang, Gan, Chen, Rho and Hong [21] proposed FUP-HUPtree-MOD algorithm to efficiently maintain and update the built HUP (High Utility Pattern) tree for transaction modification based on the FUP concept. When transactions are modified from the original database, the proposed algorithm compares the items or itemsets with four predefined cases. Each case is then processed by the designed procedure to efficiently update the already built HUP tree.

### III. PROBLEM STATEMENT AND DEFINITIONS

On-Shelf utility mining extracts itemset from temporal database. In real applications, some items or itemsets may have negative profits (free items). Traditional algorithms did not consider such negative profit items. Similarly, these algorithms considers static database. Actually, temporal databases are continually appended or updated. When new transactions are modified, some new items or itemsets are introduced as high on-shelf utility itemset or some may be removed. In this paper, an effective algorithm is proposed to solve the above limitations for dynamic updated databases.

**Definition 1.** $T = \{tr_1, tr_2, tr_3, \ldots, tr_m\}$ is a set of Transactions.

**Definition 2.** $I = \{i_1, i_2, i_3, \ldots, i_n\}$ is a set of items which is included in the transaction.

**Definition 3.** $P = \{p_1, p_2, p_3, \ldots, p_k\}$ is a set of disjoint time period. In given example, value of time period p is 3.

**Definition 3.** $D = \{Tr_{1.1}, Tr_{1.2}, Tr_{1.3}, \ldots, Tr_{i.j}\}$ is a set of transcations with their occurring time. $Tr_{i.j}$ is the $j^{th}$ transaction of $i^{th}$ time period.

**Definition 4.** Utility of an item is denoted as,

$$u(i, Tr_{i.j}) = pr(i) * q(I, Tr_{i.j}) \quad (1)$$

Where, pr(i) is the external utility of an item defined in utility table and q is the internal utility (quantity) of an item in the transaction $tr_{i.j}$.

Ex – Internal utility of an item A in TID 3 is 1 and external utility is 5.

**Definition 5.** Transaction Utility of an item is the sum of utilities of all items contained in the transaction. It is denoted as,

$$tu(Tr_{i.j}) = \sum_{i \in Trans_{i.j}} u(i, Tr_{i.j}) \quad (2)$$

Ex – $tu(Tr_{1.1})$ is 31.

**Definition 6.** Periodical utility of an item in a time period is the sum of utilities of all items in all transactions of a time period.

$$pu(i, p_i) = \sum_{Tr_{i.j} \subseteq D_i \wedge X \subseteq Tr_{i.j}} u(i, Tr_{i.j}) \quad (3)$$

Ex – pu(C, T1) is 16.

**Definition 7.** Periodical Total Transaction Utility is the sum of transaction utilities within $i_{th}$ time period $t_i$.

$$pttu(t_i) = \sum_{Tr_{i.j} \subseteq D_i} tu(Tr_{i.j}) \quad (4)$$

Ex – $pttu(T_3)$ is 108.

**Definition 8.** On-shelf utility of an item is the sum of periodical utilities of an item within the union of on-shelf time periods of an item.

$$ou(i) = \sum_{Tr_{i.j} \subseteq D_i} pu(Tr_{i.j}) \quad (5)$$

**Definition 9.** On-shelf utility ratio of an item or itemset is Ratio of sum of utilities of an item or itemset within all on-shelf time periods of item or itemset and periodical total transaction utility.

**Definition 10.** Transaction weighted utility of an item or itemset is the sum of transaction utilities all transactions which contains an item or itemset.

$$twu(i, Tr_i) = \sum_{i \subseteq Tr_{i.j} \subseteq D_i} tu(Tr_{i.j}) \quad (6)$$

Ex –twu(D, T1) is tu(t1) + tu(t3) = 31+41 = 74.

**Definition 11.** An itemset is called high on-shelf utility item if its on-shelf utility ratio is larger than or equal to minimum utility.

**Definition 12.** $Tm = \{tr_1, tr_2, tr_3,…,tr_n\}$ is a set of modified transactions.

**Definition 13.** $TWU^m(i)$ is the difference of TWU of itemset i before and after modification.

$$TWU^m(i) = TWU^{Tm}(i) - TWU^T(i). \quad (7)$$

Transaction database with its occurring time, utility and PTTU table are used for the algorithm.

TABLE I.  TRANSACTION DATATABSE WITH TIME PERIOD 3.

| Time Period | TID/Item | A | B | C | D |
|---|---|---|---|---|---|
| T1 | 1 | 5 | 4 | 2 | 1 |
| | 2 | 2 | 0 | 4 | 0 |
| | 3 | 1 | 1 | 10 | 7 |
| T2 | 4 | 4 | 7 | 3 | 0 |
| | 5 | 0 | 0 | 6 | 0 |
| | 6 | 10 | 9 | 0 | 2 |
| T3 | 7 | 2 | 5 | 4 | 1 |
| | 8 | 0 | 0 | 0 | 2 |
| | 9 | 10 | 2 | 4 | 7 |

TABLE II.  UTILITY

| Item | Utility |
|---|---|
| A | 5 |
| B | -2 |
| C | 1 |
| D | 4 |

TABLE III.  TRANSACTION UTILITY

| Time Period | TID | Transaction Utility |
|---|---|---|
| T1 | 1 | 31 |
| | 2 | 14 |
| | 3 | 43 |
| T2 | 4 | 23 |
| | 5 | 6 |
| | 6 | 58 |
| T3 | 7 | 18 |
| | 8 | 8 |
| | 9 | 82 |

TABLE IV.  PTTU

| Time Period | PTTU |
|---|---|
| 1 | 88 |
| 2 | 87 |
| 3 | 108 |

## IV. PROPOSED ALGORITHM

In this paper, an efficient algorithm **F**ast **U**pdated Utility **P**attern **T**ree for **H**igh **O**n-shelf **U**tility **I**tems with **N**egative value (FUPT-HOUIN) is proposed for finding high on-shelf utility items of both positive and negative value from **dynamic updated database.**

FUP concept is used to avoid unnecessary candidate generation and improve execution efficiency. Two phase model is used to build **utility tree**. In first phase, transaction weighted utility of all items is calculated. High transaction weighted utility 1-itemset are sorted in descending order. In second phase, tree is constructed by scanning each transaction in the database.

Header table is built for traversing the tree. The header table consists sorted high transaction weighted utility items and their pointers which link to their first occurrence nodes in the tree. Figure 1. shows an example of built utility tree.

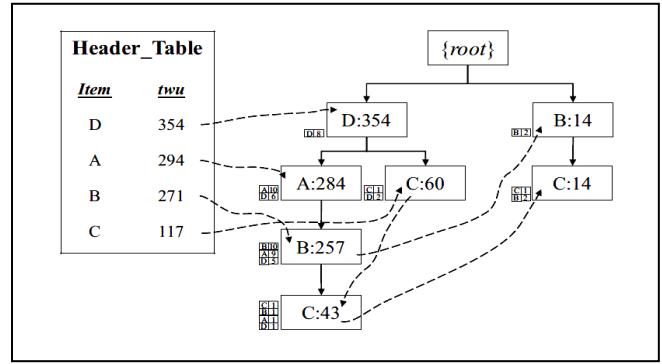

Figure 1.  Example of Utility Tree [21].

The proposed algorithm will scan the original database to build utility tree. Then it finds high on-shelf utility of an itemset. When any transaction is modified, it calculates its transaction weighted utility difference before and after modification for 1-itemset. Based on difference value, it decides the case of 1-itemset among predefined four cases and stores the itemset into insert, decrease or rescan set. Four cases are –

- Case 1: An item or itemset is frequent in an original database and modified database.
- Case 2: An item or itemset is frequent in an original database but not in the modified database.
- Case 3: An item or itemset is not frequent in an original database but may be frequent in the modified database.
- Case 4: An item or itemset is not frequent in an original database as well as modified database.

Each case is processed separately. For case 1 and 2, the header table and utility tree is updated whenever required. Rescanning of an entire database is necessary only for case 3. Flowchart of proposed algorithm is shown in Figure 2.

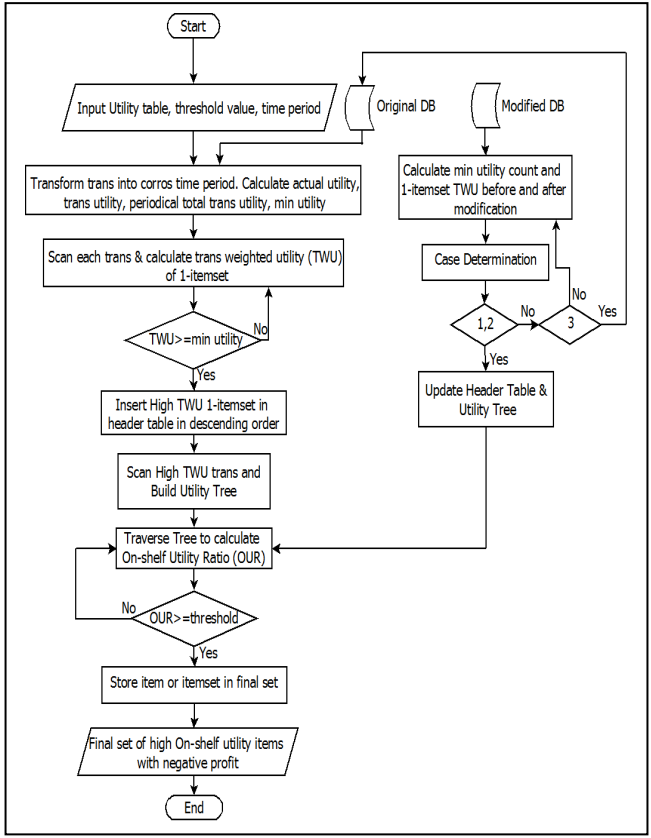

Figure 2. Flow chart of proposed algorithm.

*A. Algorithm*

Proposed algorithm is divided into two subtasks: The first is to build utility tree by scanning original database to find high on-shelf utility itemsets with negative values, while other is to update utility tree and find high on-shelf utility itemsets with negative values from dynamic updated database (Modified database).

Subtask 1: Build Utility Tree and find on-shelf itemsets.

Input: Temporal Database with transaction occurring time, User specified minimum utility threshold value, time period.
Output: Utility Tree and a set of high on-shelf utility items with negative values from original database.

STEP 1. Transform occurring time of each transaction into corresponding time period.
STEP 2. For each transaction in each time period,
    a) Calculate actual utility, transaction utility and transaction weighted utility.
    b) Calculate Periodical total transaction utility and minimum Utility.
STEP 3. Build Utility tree for each time period.
    a) Find 1-itemset with high transaction weighted utility itemsets and sort in descending order.
    b) Build header table with high transaction weighted utility itemsets.
    c) Scan database and Insert High transaction weighted utility itemset with Negative items in the tree.
STEP 4. Traverse the tree using header table for each time period.
    a) Scan database to calculate periodical utility of an itemset for time periods in which it does not appear.
    b) Calculate on-shelf utility ratio.
    c) If it is larger than or equal to threshold value, store it in final set HOUIN.

Subtask 2: Update Utility Tree and find on-shelf itemsets for modified transactions.

Input: Temporal Database with transaction occurring time, User specified minimum utility threshold value, modified database.
Output: A set of high on-shelf utility items with negative values for modified database.

STEP 1. Modified Transactions are saved in modified db.
STEP 2. For each modified transaction in modified db
    a) Find minimum utility count and transaction weighted utility difference for each 1-itemset.
    b) If an itemset's difference is larger than minimum utility count, itemset is inserted in both increase and decrease sets. Update transaction weighted utility in the header table.
    c) If an itemset's difference is smaller than minimum utility count, update child nodes of itemsets in utility tree by connecting them to the parent node of itemset and remove the itemset from tree and header table.
    d) If an itemset is not in header table and its difference is larger than minimum utility count, then it is inserted in rescan set.
STEP 3. Update utility tree by finding the corresponding branches of itemsets in increase, decrease and rescan set. For rescan set, items are sorted and inserted at the end of header table.
STEP 4. Traverse the updated tree and update the set HOUIN.

## V. EXPERIMENTAL RESULTS

In this section, performance of proposed algorithm is compared with existing algorithm HOUN for high on-shelf utility mining with negative item value.

A dataset, BMS-POS (Frequent Itemsets Mining Dataset Repository) is used in the experiments. It is downloaded

from frequent itemset repository [22]. The dataset is used in the KDDCUP 2000 competition which contains several years of point-of-sale data from a large electronics retailer. Each transaction contains many items. Quantities and utilities of items are generated by simulated model. These values are generated randomly. Experiments are performed on 5K transactions and 788 distinct items. Time period taken is 3.

As, there are many items in one transaction, chances of items occurrence in each transaction is high. Number of candidate itemsets generated decreases as the threshold value increases. Thus execution time is more for small threshold value and less for high threshold value.

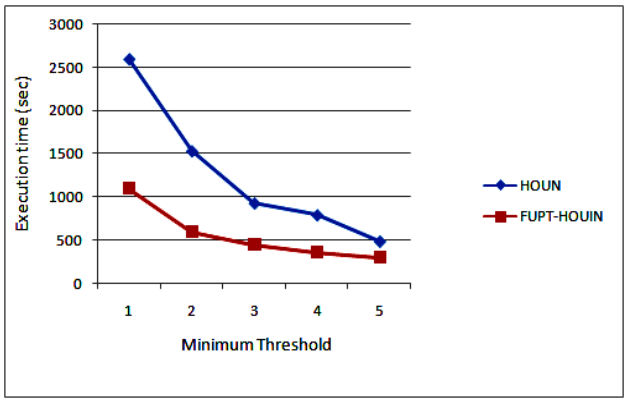

Figure 3.  Execution efficiency under various thresholds.

Figure 3. shows the execution time required for proposed and existing algorithm. Proposed algorithm requires less time than existing algorithm because it builds utility tree to avoid candidate itemset generation.

Experiments for modified transactions are performed with minimum threshold value 5. Figure 4. shows execution time required for modified transactions. In case of transaction modification, existing algorithm rescans the database. Thus whole process is repeated and increases execution time. In proposed algorithm, utility tree is updated and it does not require scanning the database every time. Database is rescanned if required. Thus it takes less time to execute than existing algorithm.

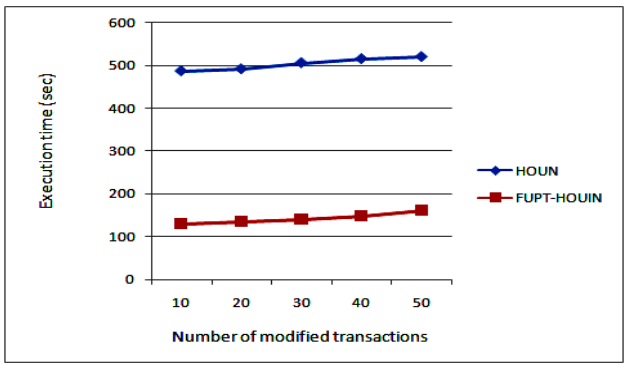

Figure 4.  Execution efficiency under various modified transactions.

## VI. CONCLUSION

In this paper, we have proposed an algorithm FUPT-HOUIN for finding high on-shelf utility itemsets with negative item values from dynamic updated database. In first phase, an algorithm builds utility tree by scanning original database. In second phase, it updates utility tree. It rescans an original database whenever necessary. Database is not rescanned for every modification which reduces the execution time of an algorithm. Whenever transactions are modified, utility tree is updated based on predefined four cases. Utility tree avoids unnecessary candidate itemset generations and thus improves execution time. Experimental results show that proposed algorithm requires less execution time than existing algorithm.

## VII. FUTURE SCOPE

In future, we can further extend our algorithm for other maintenance problem like transaction insertion, deletion. Similarly, this algorithm can be further applied to other practical applications such as data streams.

ACKNOWLEDGMENT

The authors would like to thank the researcher as well as publishers for making their resources available and the teachers for their guidance. We are thankful to authorities of Savitribai Phule Pune University for their constant guidelines and support. We also thank the college authorities for providing the required infrastructure and support. Finally, we would like to extend a heartfelt gratitude to all friends and family members.